# Three-Dimensional Wave Behaviour of Light

F. Logiurato, B. Danese, L. M. Gratton and S. Oss

Physics Department, Trento University, Italy

June 14, 2006

*Abstract*—We describe a simple experimental apparatus which allows one to observe the wave properties of light in a new way. This apparatus also makes possible to introduce and illustrate, in a very suggestive way, some fundamental principles of quantum theory.

## I. INTRODUCTION

Quantum theory is introduced in many books by means of an example widely ecognized as paradigmatic: the double-slit experiment (see, *e.g.,* Refs. 1-4). Light, after travelling and behaving as a wave, manifests itself on the detection screen as a stream of corpuscles. According to Feynman, it is *absolutely* impossible to explain this phenomenon in any classical way. In his opinion this is the "heart" of quantum mechanics, *"in reality, it contains the only mystery"* [1].

However, in experiments emphasizing the wave nature of light, diffraction and interference patterns are shown only in the last part of the light path. What is going on in the space between the slits and the detection screen is only sketched in the figures. We developed a simple apparatus where diffraction and interference patterns are not only observed at the final screen position as in traditional experiments, but also in a three-dimensional environment [5, 6]. In this paper we give a few examples of how our apparatus may be used to illustrate the wave properties of light.

## II. EXPERIMENTAL SETUP AND RESULTS

Many simple techniques have been adopted in the past to visualize light rays. For instance, light diffusion from chalk powder or from smoke. The technique we adopt here is based on light diffusion from water droplets produced by an ultrasonic mist-maker immersed in water. Vibrations at ultrasonic frequencies of a ceramic electrode inside the mist-maker generate ultrasounds that break the surface of the liquid and nebulize the water. This technique produces a continuous and homogeneous fog which allows the formation through its whole volume of very stable luminous patterns.

To minimize turbulences and to assure high homogeneity of the fog along the light path, the mist-maker is placed in a box with transparent walls, such as an aquarium. A black piece of fabric covers the walls of the box through which no vision takes place, to avoid disturbing reflections.

As coherent light source we use a 10 mW HeNe laser. The laser wavelength is $\lambda = 0.6328$ μm. The slits belong to Pasco optical kit OS-9165. The photographs are taken with a digital D70 reflex camera. The equivalent sensitivity is set to 200 ISO. Exposure times ranges from $1/30^{th}$ to *1/2* sec, with various f/values.

In Figure 1 we see various images. In each of them HeNe laser light impinges on a single slit, and the slits in different images have different widths. The dependence of the extent of the diffracted beam on the width of the slit is clear: the narrower the slit the broader the intense ($0^{th}$ order) central beam.

We think that these images are a really beautiful illustration of the famous experiment of the single slit by which Heisenberg introduces the uncertainty relations [7, 8]. In fact, if we regard light as a stream of corpuscles (the photons), it follows that the narrower the slit, the higher the space localization of each photon in the light beam, and the larger the uncertainty of the momentum acquired by it.

In Figure 2 two further interesting images are compared. In the left one, the light beam impinges on a screen with two slits. The beam that passes through the screen forms the well-known interference fringes of the classic Young experiment in the space. This experiment provided the definitive demonstration of the existence of wave properties of light in 1802. In the resulting series of maxima and minima we can distinguish two patterns: the enveloping pattern due to the light diffraction through each slit and, inside the envelope, the interference pattern of the light coming from the two slits [9]. We may compare the Young experiment of the double slit with the diffraction from the single slit (right image). In the latter, the slit has the same width as each slit in the Young experiment. It can be noted immediately that the interference pattern from two slits is not the sum of two diffraction patterns from single slits. This phenomenon cannot be explained if one adopts only the classic corpuscular model of light. Hence, the photos in Figure 2 support effectively the undulatory counterpart of Feynman's two-slits experiment, by which this author introduces the wave-corpuscle dualism and Bohr's complementarity principle [1].

fabrizio.logiurato@unitn.it

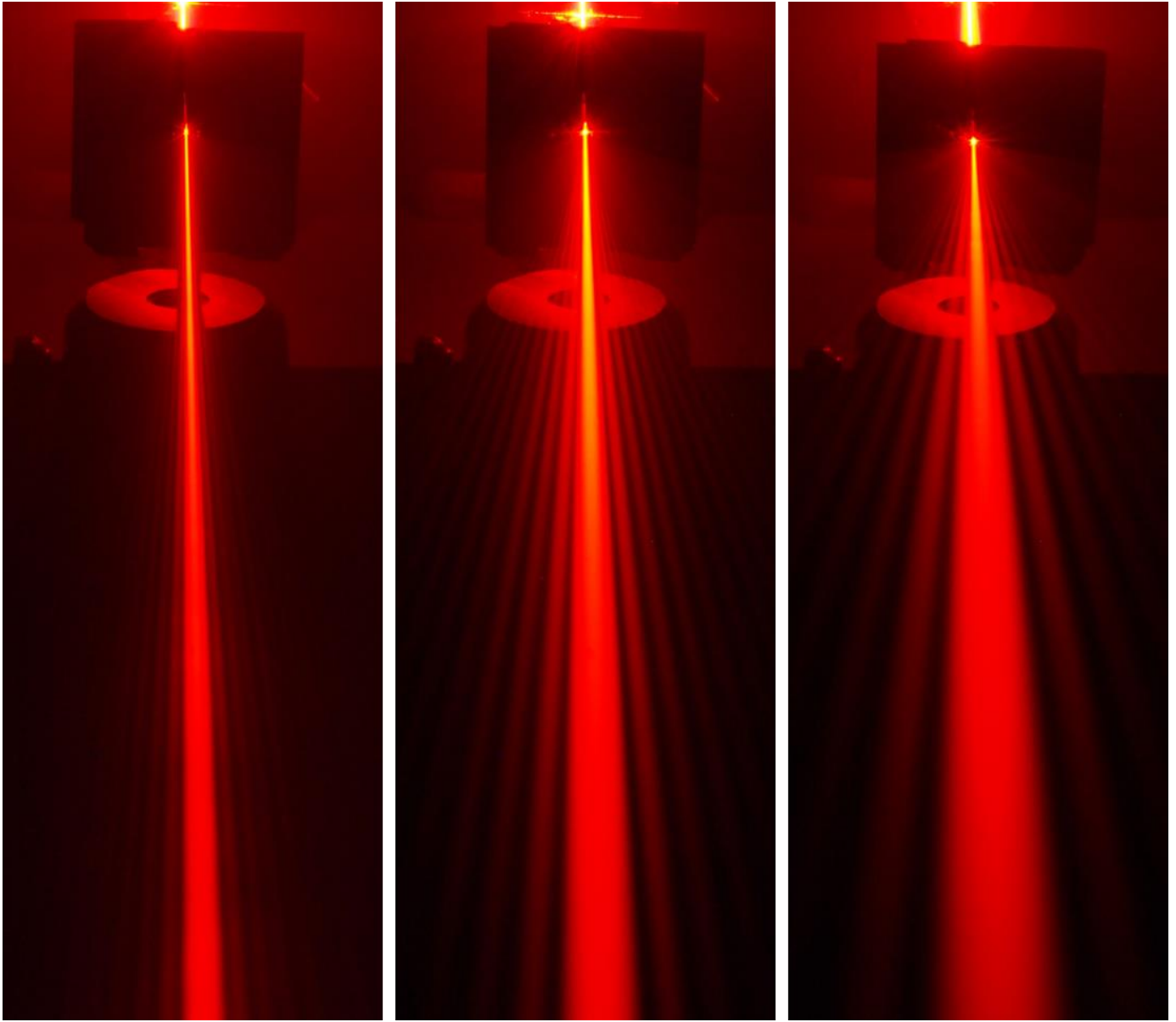

Figure 1. The experiments of diffraction from a single slit. The images correspond to slits of decreasing width (from left to right): 80 µm, 40 µm, 20 µm.

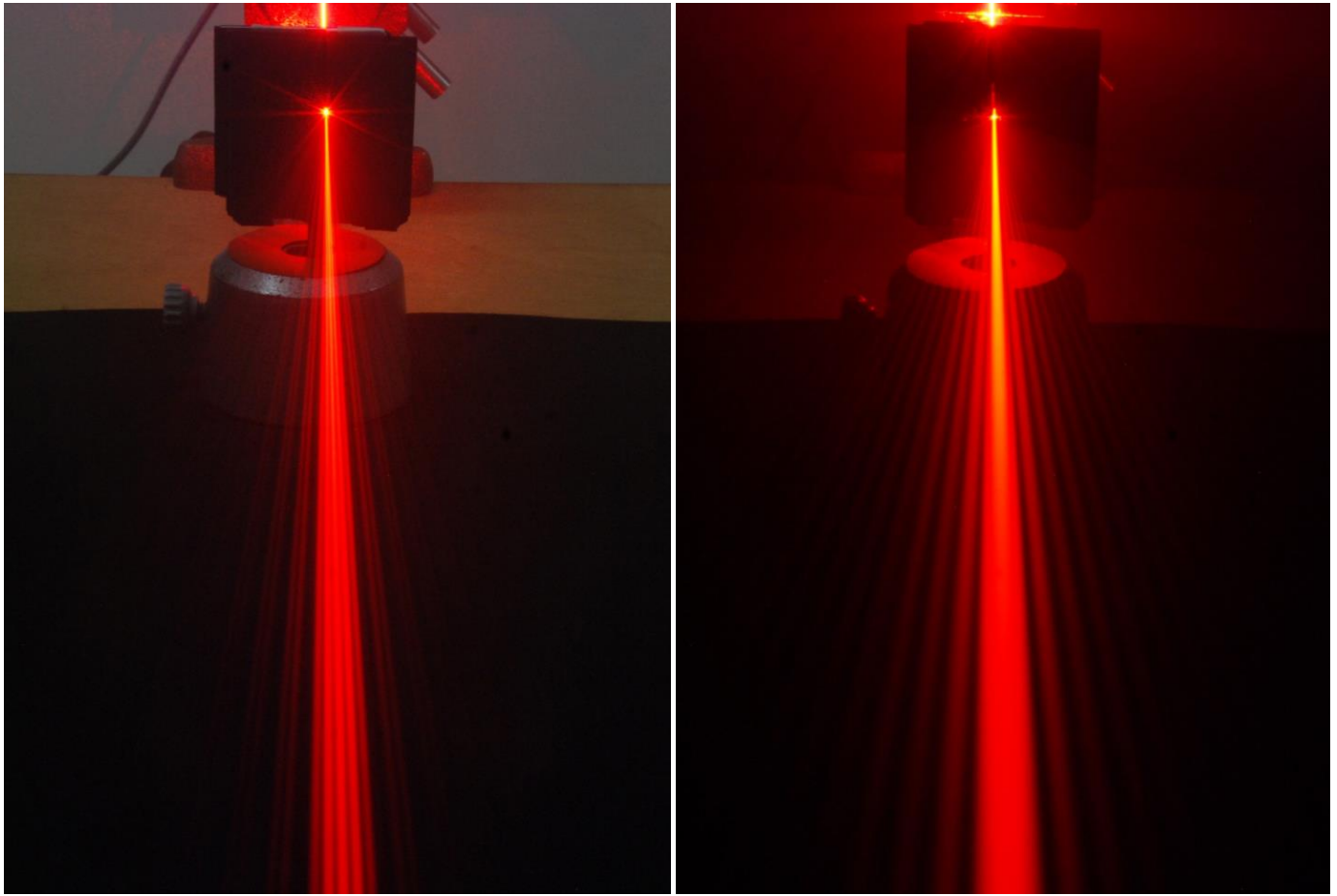

Figure 2. Left: Young experiment of two slits. The width of the slits is 40 µm, and their separation is 125 µm. Right: diffraction from a single slit of the same width.